\documentclass[preprint,aps,nofootinbib]{revtex4}
\usepackage[dvips]{graphicx}
\usepackage{amssymb}
\usepackage{amsmath}
\usepackage{amsmath, amsthm, amssymb, mathrsfs}
\usepackage{pstricks}

\newcommand{\ket}[1]{\lvert #1 \rangle}
\newcommand{\bra}[1]{\langle #1 \lvert}

\begin{document}

\title{Quantum Resonance near Optimal Eavesdropping in Quantum Cryptography} 

\author{
Eylee Jung$^{1}$, Mi-Ra Hwang$^{1}$, DaeKil Park$^{1}$, Hungsoo Kim$^{2}$,
Jin-Woo Son$^{3}$, Eui-Soon Yim$^{4}$, Seong-Keuck Cha$^{5}$, S. Tamaryan$^{6}$,
Sahng-Kyoon Yoo$^{7}$}

\vspace{1.0cm}

\affiliation{
$^1$ Department of Physics, Kyungnam University, Masan, 631-701, 
Korea  \\
$2$ Department of Applied Mathematics, Pukyong National University, Pusan, 606-737,
Korea  \\
$^3$ Department of Mathematics, Kyungnam University, Masan,
631-701, Korea         \\
$^4$ Department of Computer Science, Semyung University, Chechon, 390-711, Korea  \\ 
$^5$ Department of Chemistry, Kyungnam University, Masan, 631-701, Korea   \\
$^6$ Theory Department, Yerevan Physics Institute,
Yerevan-36, 375036, Armenia   \\
$^7$ Green University, Hamyang, 676-872, Korea }

\vspace{1.0cm}

\begin{abstract}

We find a resonance behavior in the disturbance when an eavesdropper chooses a near-optimal
strategy intentionally or unintentionally when the usual Bennett-Brassard cryptographic scheme
is performed between two trusted parties. This phenomenon tends to disappear when eavesdropping
strategy moves far from the optimal one. Therefore, we conjecture that this resonant effect is 
a characteristic for the eavesdropping strategy near to optimal one.
We argue that this effect makes the quantum cryptography
more secure against the eavesdropper's attack.

\end{abstract}


\maketitle

Recently, there are a lot of activities in the various applications of the quantum information 
theories\cite{nielsen00}. Among them the most important ones are application to quantum computer
and quantum cryptography.
While the physical realization of the quantum computer seems to be 
far from the embodiment in a few years, 
developing quantum cryptography based on the BB84\cite{bb84} and 
Ekert91\cite{ek91} protocols is at the stage of the industrial era\cite{all07}. 

The main issue of the quantum cryptography is to determine how secure the quantum cryptographic
scheme compared to the classical scheme. This issue can be turned into the following 
question: how much information the eavesdropper (Eve) can gain when a secret key is 
established between two trusted parties(Alice and Bob)? 

Of course, the answer of the question is dependent on the eavesdropping strategies. The authors
in Ref.\cite{fuchs97,griff97} computed the optimal mutual information between Alice and Eve
in the usual BB84 protocol and the final results are
\begin{equation}
\label{optimal-1}
I_{xy} = \frac{1}{2} \phi \left[2 \sqrt{D_{uv} (1 - D_{uv})} \right]  \hspace{1.0cm}
I_{uv} = \frac{1}{2} \phi \left[2 \sqrt{D_{xy} (1 - D_{xy})} \right],
\end{equation}
where $\phi(z) = (1+z) \log_2 (1+z) + (1-z) \log_2 (1-z)$ and the subscripts denote the conjugate
basis Alice and Bob use during BB84 process. The disturbance $D$ is Bob's observable error rate.

Subsequently, the BB84 protocol has been extended to the case that Alice and Bob use the three
conjugate bases\cite{bruss98}. It has been shown that this extended scheme is more secure against
the optimal eavesdropping. In order to find more secure quantum cryptographic protocols, recently,
much attention is paid to the qutrit\cite{bech00,bruss02}, qudit\cite{bruss02,boure01,cerf01} and
continuous-variable systems\cite{piran08}. The optimal eavesdropping on noisy states is also
fully discussed very recently in Ref.\cite{shad08}.

Instead of the optimal eavesdropping strategy we would like to discuss, in this letter, on the 
near-optimal eavesdropping in usual BB84 scenario. We will show that an interesting quantum 
resonance occurs in the disturbance between Alice and Bob when Eve's eavesdropping strategy is 
near to optimal.

First, we consider a simple case that Eve uses one-qubit probe.
Eve makes contact her probe with the qubit between Alice and Bob and gets her probe
to be entangled. We restrict ourselves into the case when Alice chooses $x-y$ basis with notation
$\ket{x} \equiv \ket{0}$ and $\ket{y} \equiv \ket{1}$. We choose the entangled states as following
\begin{equation}
   \begin{array}{l}
      \ket{x} \rightarrow \ket{X}
	     =
		    a \ket{00} + b \ket{11}
			                                                                                             \\
\ket{y} \rightarrow \ket{Y}
		   =
		      \delta
			  \left(
			             -b \ket{00} + a \ket{11}
			  \right)
			+\sqrt{1 - \delta^2}
			  \left(
			          c \ket{10} + d \ket{01}
			  \right)
   \end{array}
\end{equation}
with $a^2 + b^2 = c^2 + d^2 = 1$.

According to BB84 scenario, Alice will announce bases which she used to establish a 
secret key through public channel. After the announcement Eve performs an appropriate 
quantum-mechanical measurement on her probe to gain information on the Alice's qubit. 
In order to maximize the information gain one can show that Eve takes a POVM measurement with
complete set of positive operators $\{ E_0 = \ket{E_0}\bra{E_0}, E_1=\ket{E_1}\bra{E_1} \}$,
where\cite{fuchs96}
\begin{eqnarray}
\label{povm1}
& &\ket{E_0} = -\frac{1}{\sqrt{2}} \epsilon(ac - bd) \sqrt{1 + \cos \varphi} \ket{0} + 
\frac{1}{\sqrt{2}} \sqrt{1 - \cos \varphi} \ket{1}    \\    \nonumber
& &\ket{E_1} = \frac{1}{\sqrt{2}} \sqrt{1 - \cos \varphi} \ket{0} + \frac{1}{\sqrt{2}}
\epsilon(ac - bd) \sqrt{1 + \cos \varphi} \ket{1}.
\end{eqnarray}
In Eq.(\ref{povm1}) $\epsilon(x) = x / |x|$ is usual alternating function and 
\begin{equation}
\label{povm2}
\cos \varphi = \frac{\alpha}{\sqrt{\alpha^2 + \beta^2}}
\end{equation}
with $\alpha = (a^2 - c^2) - \delta^2 (b^2 - c^2)$ and 
$\beta = \delta \sqrt{1 - \delta^2} (ac - bd)$. Then, following Ref. \cite{fuchs97}, it is 
straightforward to compute the Eve's average information gain $G$:
\begin{equation}
\label{gain1}
G = q_0 G_0 + q_1 G_1
\end{equation}
where
\begin{eqnarray}
\label{gain2}
& & q_0 = \frac{1}{2} + \frac{1}{4} (1 - \delta^2) (a^2 - b^2 + c^2 - d^2) \cos \varphi 
- \frac{1}{2} \delta \sqrt{1 - \delta^2} |ac - bd| \sin \varphi     \\   \nonumber
& & q_1 = \frac{1}{2} - \frac{1}{4} (1 - \delta^2) (a^2 - b^2 + c^2 - d^2) \cos \varphi 
+ \frac{1}{2} \delta \sqrt{1 - \delta^2} |ac - bd| \sin \varphi     \\   \nonumber
& &G_0 = \frac{1}{4 q_0}
\bigg| \left\{ (a^2 - b^2 - c^2 + d^2) + \delta^2 (a^2 - b^2 + c^2 - d^2) \right\} \cos \varphi
+ 2 \delta \sqrt{1 - \delta^2} |ac - bd| \sin \varphi \bigg|      \\    \nonumber
& &G_1 = \frac{1}{4 q_1}
\bigg| \left\{ (a^2 - b^2 - c^2 + d^2) + \delta^2 (a^2 - b^2 + c^2 - d^2) \right\} \cos \varphi
+ 2 \delta \sqrt{1 - \delta^2} |ac - bd| \sin \varphi \bigg|.
\end{eqnarray}
Using Eq.(\ref{gain2}) one can compute the mutual information between Alice and Eve, which is 
\begin{equation}
\label{mutual1}
{\cal I}_{AE} = \frac{1}{2} \left[ q_0 \phi(G_0) + q_1 \phi(G_1) \right]
\end{equation}
where $\phi(z) = (1+z) \log_2 (1+z) + (1-z) \log_2 (1-z)$.

Now, let us discuss on the Bob's error rate $D_B$, which is usually called disturbance.
Firstly, let us consider $d_{\lambda u}$ (or $d_{\lambda v}$), which is the probability
that Bob gets a wrong result {\it conditioned upon} Alice sending $\ket{u}$ (or $\ket{v}$)
and Eve measuring $\lambda$, where $\ket{u} = (\ket{x} + \ket{y})/\sqrt{2}$ and 
$\ket{v} = (\ket{x} - \ket{y})/\sqrt{2}$. Then, it is easy to show 
\begin{equation}
\label{disturbance1}
d_{\lambda u} = 1 - \frac{\bra{U} (\ket{u}\bra{u}) \otimes E_{\lambda} \ket{U}}
                         {\bra{U} \openone \otimes E_{\lambda} \ket{U}}
\hspace{1.0cm}
d_{\lambda v} = 1 - \frac{\bra{V} (\ket{v}\bra{v}) \otimes E_{\lambda} \ket{V}}
                         {\bra{V} \openone \otimes E_{\lambda} \ket{V}}
\end{equation}
where $\ket{U} = (\ket{X} + \ket{Y}) / \sqrt{2}$ and $\ket{V} = (\ket{X} - \ket{Y}) / \sqrt{2}$.
If Eve has chosen the optimal strategy {\it ab initio}, $d_{\lambda u}$ and $d_{\lambda v}$ 
should coincide with each other. Since, however, we are considering the non-optimal case,
we cannot expect $d_{\lambda u} = d_{\lambda v}$ in general. Although it is straightforward
to compute $d_{\lambda u}$ and $d_{\lambda v}$, we will not present the explicit expressions 
in this letter due to their lengthy expressions. As expected, $d_{\lambda u}$ is different
from $d_{\lambda v}$ except $\delta = 0$ case. 

\begin{figure}[ht!]
\begin{center}
\includegraphics[height=8.0cm]{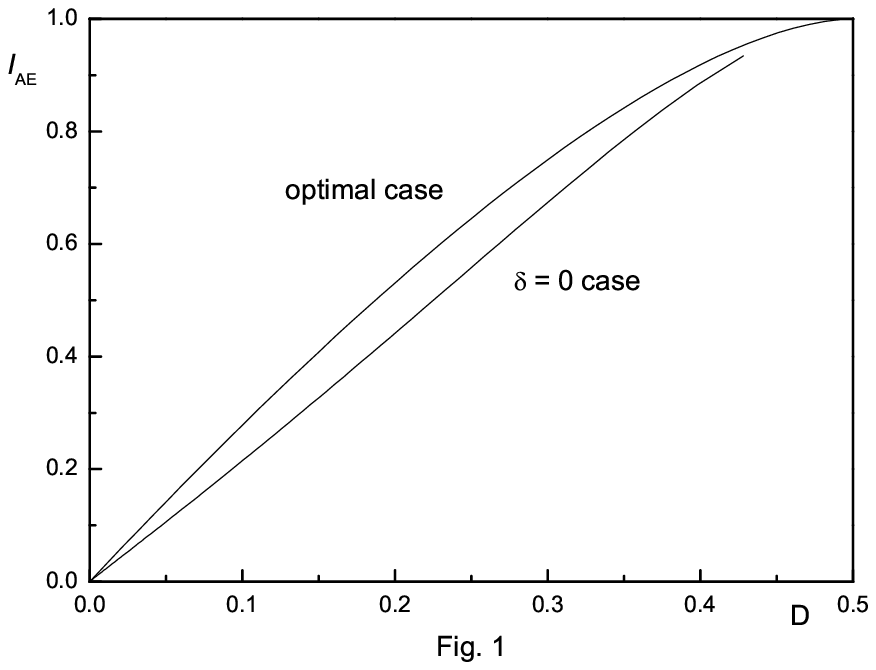}
\caption[fig1]{ 
The plot of $D$-dependence of the mutual information between Alice and Eve when $\delta = 0$ and 
$a = 0.01$. Because of the fact that $d_{\lambda}$ is dependent on $\lambda$, the mutual 
information for the $\delta = 0$ case is slightly smaller than that for the optimal case.}
\end{center}
\end{figure}

We would like to discuss the $\delta = 0$ case briefly. In this case the most optimal conditions
derived in Ref.\cite{fuchs97} are satisfied. The only one this case does not satisfy is the
fact that $d_{\lambda} \equiv d_{\lambda u} = d_{\lambda v}$ is dependent on $\lambda$. This fact
makes the mutual information ${\cal I}_{AE} (\delta = 0)$ to be slightly smaller than the 
optimal value Eq.(\ref{optimal-1}) as shown in Fig. 1. In Fig. 1 the disturbance $D$ is defined
as an average Bob's error rate
\begin{equation}
\label{disturbance-2}
D = \sum_{\lambda} q_{\lambda} d_{\lambda} = \frac{1}{2} (1 - ac - bd).
\end{equation}
Since the $\delta = 0$ case does satisfy the almost optimal conditions, we guess that the mutual 
information ${\cal I}_{AE}$ for this case is maximum on condition that Eve uses the 
single-qubit probe.

Now let us consider the $\delta \neq 0$ case. Since, in this case, $d_{\lambda u}$ is different
from $d_{\lambda v}$, we should define the disturbance as 
\begin{equation}
\label{disturbance-3}
D = p_{u} D_u + p_v D_v
\end{equation}
where $p_i$'s are the prior probabilities that Alice sends signal $i$, and 
\begin{equation}
\label{disturbance-4}
D_u \equiv \sum_{\lambda} q_{\lambda} d_{\lambda u}    \hspace{1.0cm}
D_v \equiv \sum_{\lambda} q_{\lambda} d_{\lambda v}.
\end{equation}
In this letter we take a reasonable assumption that the two signals are equiprobable, i.e.
$p_u = p_v = 1/2$.

\begin{figure}[ht!]
\begin{center}
\includegraphics[height=8.0cm]{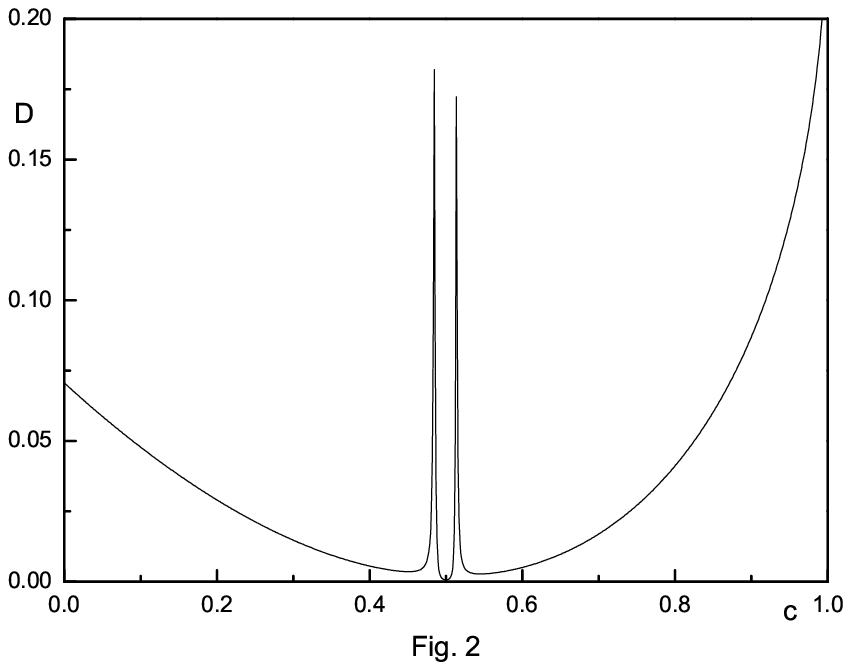}
\caption[fig2]{The $c$-dependence of $D$ when $\delta = 0.05$ and $a = 0.5$. Fig. 2 implies that 
there exists a resonance-like phenomenon in the Bob's error rate when Eve chooses the near-optimal
eavesdropping strategy. The two peaks in the figure are originated from $D_u$ and $D_v$ 
respectively.}               
\end{center}
\end{figure}

Fig. 2 is a $c$-dependence of $D$ when $\delta = 0.05$ and $a=0.5$. Fig. 2 shows that there exist
two sharp peaks, which looks like a resonance phenomenon. The left and right peaks are 
originated from $D_v$ and $D_u$ respectively. The reason why the peaks appear in the disturbance
can be explained as follows. Under some circumstances the numerator 
$\bra{U} \left(\ket{u}\bra{u} \otimes E_{\lambda} \right) \ket{U}$ is slightly smaller than
the denominator $\bra{U} \openone \otimes E_{\lambda} \ket{U}$ in $d_{\lambda u}$ in the 
wide range of parameter space. Thus, $d_{\lambda u}$ becomes very small in this region. If 
however, there are some points where the numerator 
$\bra{U} \left(\ket{u}\bra{u} \otimes E_{\lambda} \right) \ket{U}$ approaches zero, this makes
a sharp increase at these points even if the denominator 
$\bra{U} \openone \otimes E_{\lambda} \ket{U}$ is slightly larger than the numerator. Similar
phenomenon can occurs for $d_{\lambda v}$, which gives different peak. 

Numerical calculation shows that these sharp peaks disappear when $\delta$ increases. This 
fact makes us to conjecture that this resonance-like phenomenon happens in the near-optimal
strategy because the $\delta = 0$ case can play a role as an optimal strategy 
on condition that Eve uses a single-qubit probe. 

In order to check the validity of our conjecture let us consider the case that Eve chooses the
near-optimal strategy with her two-qubit probe. We assume that entanglement between Alice's 
and Eve's qubits is given by
\begin{eqnarray}
\label{near-opt-1}
& &\ket{X} = \sqrt{s} \ket{x} \ket{\xi_x} + \sqrt{1-s} \ket{y}\ket{\zeta_x}  \\   \nonumber
& &\ket{Y} = \sqrt{s} \ket{y} \left[\sqrt{1 - \delta^2} \ket{\xi_y} + \delta
             \left( -\sqrt{1-\beta} \ket{\Psi_{xy}^+} + \sqrt{\beta} \ket{\Psi_{xy}^-} \right) 
                                                                            \right]
                                                                              \\    \nonumber
& &\hspace{2.0cm} +
             \sqrt{1-s} \ket{x} \left[\sqrt{1 - \delta^2} \ket{\zeta_y} + \delta
             \left( -\sqrt{1-\alpha} \ket{\Phi_{xy}^+} + \sqrt{\alpha} \ket{\Phi_{xy}^-} \right) 
                                                                            \right]
\end{eqnarray}
where
\begin{eqnarray}
\label{near-opt-2}
& &\ket{\xi_x} = \sqrt{\alpha} \ket{\Phi_{xy}^+} + \sqrt{1-\alpha} \ket{\Phi_{xy}^-}
\hspace{1.0cm}
\ket{\xi_y} = \sqrt{\alpha} \ket{\Phi_{xy}^+} - \sqrt{1-\alpha} \ket{\Phi_{xy}^-}
                                                                             \\   \nonumber
& &\ket{\zeta_x} = \sqrt{\beta} \ket{\Psi_{xy}^+} - \sqrt{1-\beta} \ket{\Psi_{xy}^-}
\hspace{1.0cm}
\ket{\zeta_y} = \sqrt{\beta} \ket{\Psi_{xy}^+} + \sqrt{1-\beta} \ket{\Psi_{xy}^-}.
\end{eqnarray}
The states $\ket{\Phi_{xy}^{\pm}}$ and $\ket{\Psi_{xy}^{\pm}}$ denote the maximally entangled
Bell basis as follows:
\begin{equation}
\label{bell-b}
\ket{\Phi_{xy}^{\pm}} = \frac{1}{\sqrt{2}} \left(\ket{x}\ket{x} \pm \ket{y}\ket{y} \right)
\hspace{1.0cm}
\ket{\Psi_{xy}^{\pm}} = \frac{1}{\sqrt{2}} \left(\ket{x}\ket{y} \pm \ket{y}\ket{x} \right).
\end{equation}
The reason why we choose Eq.(\ref{near-opt-1}) is that the entangled states $\ket{X}$ and 
$\ket{Y}$ with $\delta = 0$ provides an optimal mutual information to Eve as shown in 
Ref.\cite{fuchs97}. Thus, we want to find a resonance phenomenon when $\delta$ is small
to check the validity of our guess.

\begin{figure}[ht!]
\begin{center}
\includegraphics[height=8.0cm]{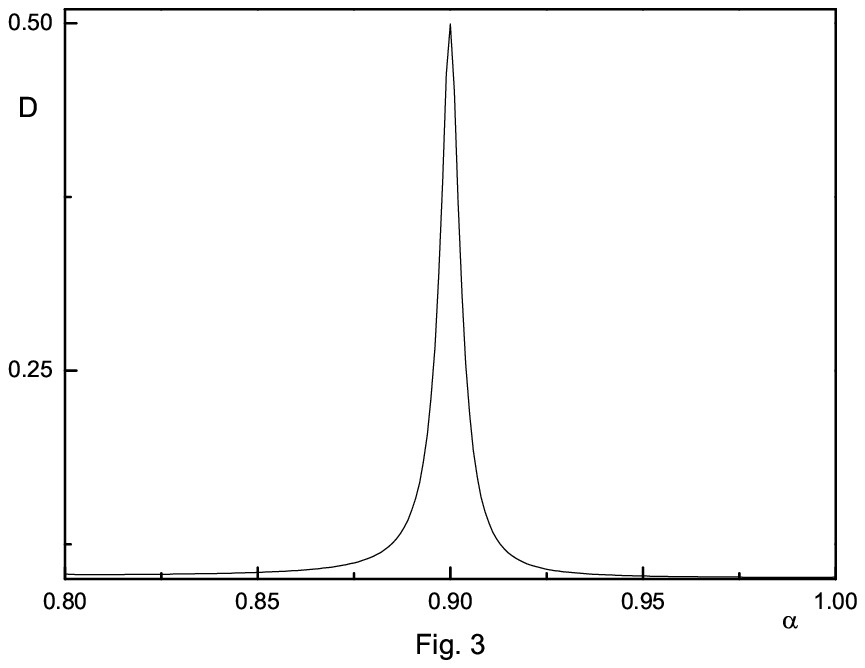}
\caption[fig3]{The $\alpha$-dependence of the disturbance $D$ when Eve uses the entanglement
Eq.(\ref{near-opt-1}). The other constants are fixed by $\beta = 1.8 - \alpha$, $s=0.5$ and 
$\delta = 0.05$. As expected the disturbance $D$ exhibits a sharp resonance. This effect
disappears with increasing $\delta$, which means that Eve's eavesdropping strategy is far
from optimal one. Thus, this resonance seems to occur in the near-optimal strategy.}
\end{center}
\end{figure}

The disturbance $D$ can be computed numerically by making use of the symbolic calculation. Fig. 3
is a plot of $\alpha$-dependence of $D$ when $\beta = 1.8 - \alpha$, $s=0.5$ and $\delta = 0.05$.
As expected the disturbance $D$ exhibits a resonance behavior with varying $\alpha$. This 
phenomenon tends to disappear with increasing $\delta$. Thus, this resonant behavior seems to be 
a characteristic for the near-optimal eavesdropping strategy. Unlike Fig. 2, Fig. 3 shows one
peak. This is due to the fact that $D_u$ and $D_v$ have peaks at the same point.

In this letter we report on the resonance phenomenon in the disturbance when Eve chooses the 
near-optimal eavesdropping strategy. In reality eavesdropper cannot perform the exact optimal
strategy due to the various nature's non-linear and/or decoherence effects. If eavesdropper
takes an near-optimal strategy, this resonance effect increases a possibility for the two
trusted parties to realize the eavesdropping attack. As a result, the resonance discussed in
this letter makes the quantum cryptography more and more secure. It is of highly important,
in this reason, to verify this resonance phenomenon in the quantum cryptographic experiment.

{\bf Acknowledgement}: 
This work was supported by the Kyungnam University
Foundation Grant, 2008.

\end{document}